\documentclass{article}

\usepackage{arxiv}

\usepackage[utf8]{inputenc} 
\usepackage[T1]{fontenc}    
\usepackage{hyperref}       
\usepackage{url}            
\usepackage{booktabs}       
\usepackage{amsfonts}       
\usepackage{nicefrac}       
\usepackage{microtype}      
\usepackage{lipsum}
\usepackage{subcaption}
\usepackage{graphicx}
\usepackage{amsmath}
\usepackage{natbib}

\usepackage{cleveref}
\usepackage{footnote}
\usepackage{tablefootnote}
\title{Artificial Buildings: Safety, Complexity and a Quantifiable Measure of Beauty}

\author{
 Arash Mehrjou\thanks{Department of Computer Science, ETH Z\"urich \& Max Planck Institute for Intelligent Systems:~\href{mailto:amehrjou@ethz.ch}{amehrjou@ethz.ch}}
}

\begin{document}
\maketitle

\begin{abstract}
A place to live is one of the most crucial necessities for all living organisms since the advent of life on planet Earth. The nature of homes has changed considerably over time. At the very early stages, human begins lived in natural places such as caves. Later on, they started to use their intelligence to build places with special purposes. Nowadays, modern technologies such as robotics and artificial intelligence have made their ways into the construction process and opened up a whole new area of opportunities and concerns that may be of interest to both technologists and philosophers. In this article, I review the evolution of buildings from fully natural to fully artificial and discuss philosophical thoughts that a fully automated construction technology may raise. I elaborate on the safety concerns of a fully automated architectural process. Then, I'll borrow Kolmogorov complexity from algorithmic information theory to define a complexity measure for buildings. The proposed measure is then used to provide a quantifiable measure of beauty.
\end{abstract}


\section{Introduction}
The concept of living organisms often comes with a property that determines a place where an organism lives. In this article, my focus will be on the organisms known as human beings since the very early years of their appearance on Earth. I use the general term ``people'' for all human beings over the history of their existence on Earth and the general term ``building'' for their living places. This article concerns one aspect of buildings which I call ``Naturalness''. Even though this term may have intuitive meaning, the exact definition requires further elaboration. It is intuitively clear that buildings were more natural initially and turned into more artificial products as a result of advances in civilization. In this article, first, the historical evolution of buildings is presented with a focus on the naturalness aspect of them. Then, the use of modern automated technologies in the construction process is introduced followed by philosophical implications that the combination of architecture and automation can bring up.

\section{From Natural to Artificial Buildings}
\label{sec:architecture_history}
Looking for \emph{natural} buildings, one can refer to caves as one of the first residential places which were found not built by humans. One example of such caves is Lascaux cave in Montignac France. This cave can be seen as a natural building that is supposedly a place of the congregation which is also decorated by ancient paintings. Regardless of the purpose of this place, it is important to notice that this place was found not made by men or their skills. Even though caves are obvious examples of natural buildings, it is difficult to define natural buildings in contemporary understanding of the construction technology. Hence, a deeper investigation into the concepts of natural and artificial buildings is required.

It is useful to agree on the definition of~\emph{nature} in the first place. Ralph Waldo Emerson (May 25, 1803 – April 27, 1882), an American philosopher published an essay on ``Nature'' in which he stated ``Whatever is not me is considered as nature''~\citep{emerson1940nature}. There is also a common understanding that ``The nature is what is untouched by men''. Emerson provided an example of a house and the role of ``will'' in its construction. The ``will'' can be seen as the design intention of the architect which is in contrast with the unconscious intent of the nature that constructs places such as caves.

A possible historical next step after caves is \emph{sod houses} which were used in various places such as in Nebraska, the U.S. in 18's. In sod houses, the grass and the soil beneath it which is held together by the grass's roots are used to construct walls. Even though these houses are built by humans, given our current idea of buildings, these houses can be considered as natural constructs since they were built with natural untouched materials.

\addtocounter{footnote}{-1}
\begin{figure}
    \begin{subfigure}[b]{.5\linewidth}
    \centering
    \includegraphics[height=8cm]{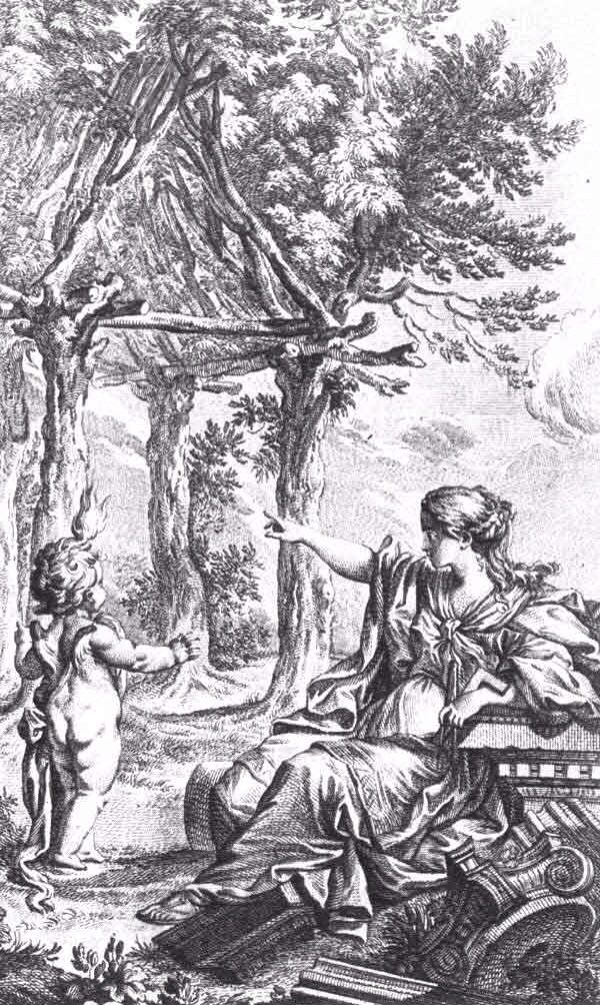}
    \caption{Charles-Dominique-Joseph Eisen's drawing for Marc-Antoine Laugier's Essai sur l'architecture\protect\footnotemark}
    \end{subfigure}%
    \begin{subfigure}[b]{.5\linewidth}
    \centering
    \includegraphics[height=8cm]{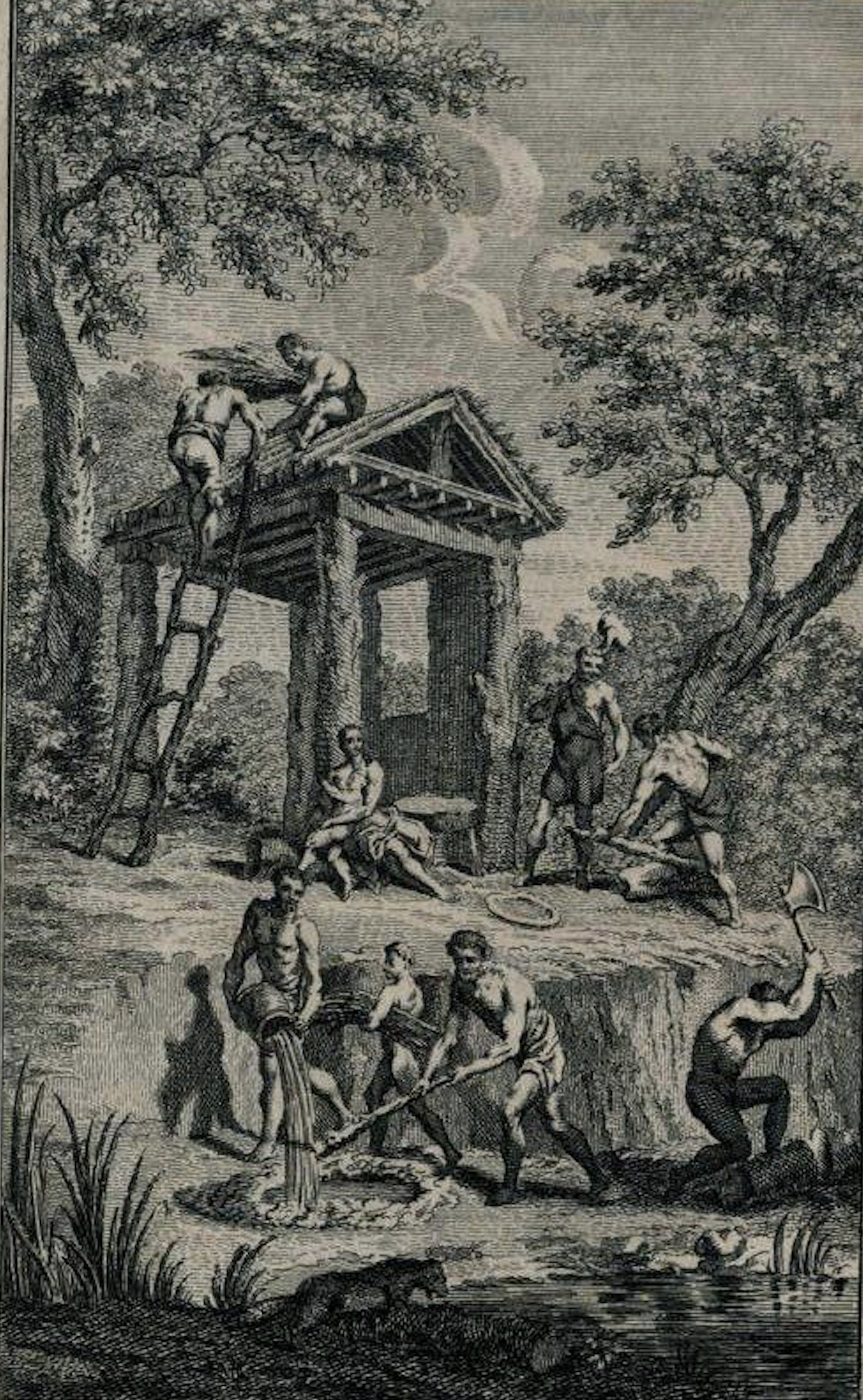}
    \caption{The frontispiece of  Marc-Antoine Laugier's Essai sur l'architecture\protect\footnotemark}
    \end{subfigure}
    \caption{Illustrations of early examples of natural buildings}\label{fig:illustrations}
\end{figure}

\addtocounter{footnote}{-1} 

\footnotetext{Photo from~\url{https://www.drawingmatter.org/sets/drawing-week/}}
\stepcounter{footnote}
\footnotetext{Photo from~\url{https://www.drawingmatter.org/sets/drawing-week/origins-translation/}}

The tension between the concepts of natural and artificial buildings can be traced back to artistic illustrations, for instance, in the paintings by Charles-Dominique-Joseph Eisen (17 August 1720 – 4 January 1778) for Marc-Antoine Laugier (January 22, 1713 – April 5, 1769) who is considered as the first architectural philosopher -- see~\Cref{fig:illustrations}. Those paintings illustrate an architect as one who willfully changes trees and other natural objects to consciously build something like a roof. Although those paintings show primitive man-made constructs, they are still inspired by nature that shows a smooth transition from natural buildings to the modern concept of artificial buildings. Another interesting idea in line with the interaction between a building and nature was \emph{Tree Shaping} proposed by the German landscape architect Arthur Wiechula (January 20, 1867 – 1941) who guided the branches of a tree to form a target design such as a building or a dance stage.

\begin{figure}[t!]
    \begin{subfigure}[b]{.5\linewidth}
    \centering
    \includegraphics[height=7cm]{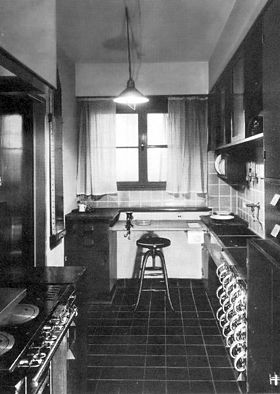}
    \caption{Entrance view}
    \end{subfigure}%
    \begin{subfigure}[b]{.5\linewidth}
    \centering
    \includegraphics[height=7cm]{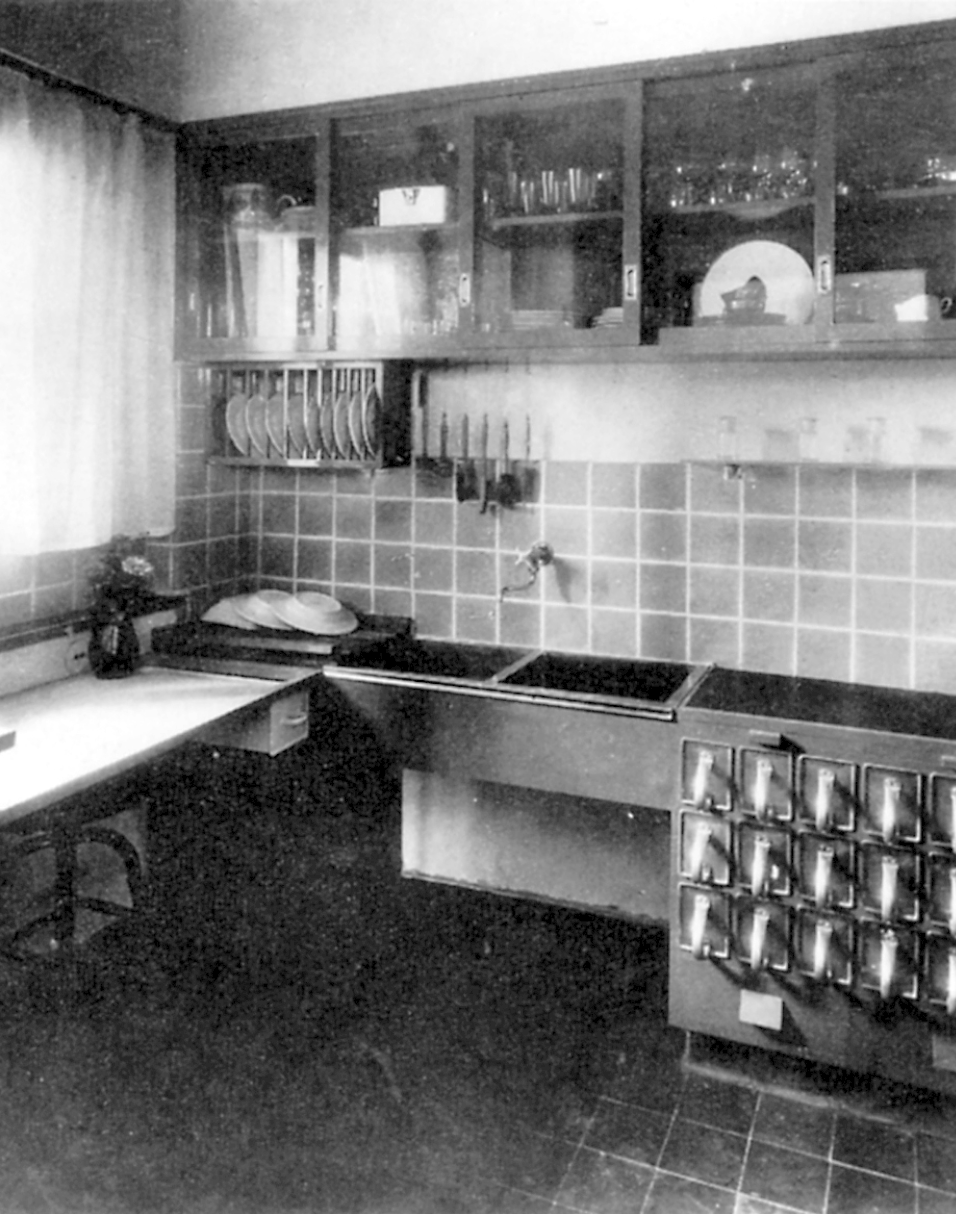}
    \caption{Inside view}
    \end{subfigure}
    \caption{Two view of the kitchen designed by Margarete Sch\"utte-Lihotzky for architect Ernst May's social housing project ``New Frankfurt'' in Frankfurt\protect\footnotemark}\label{fig:kitchens}
\end{figure}

In the early $20^\mathrm{th}$ century, industrialization took off and started to impact architecture. The new doctrine followed the logic of efficient satisfaction of a goal, for example, the rational way that a meal is prepared in a kitchen efficiently as can be seen in a couple of Margarete Schütte-Lihotzky's works (January 23, 1897, January 18, 2000) -- see~\Cref{fig:kitchens}. Some ideas of that period especially in Hugo Häring's (11 May 1882 – 17 May 1958) works, were inspired by the human body and called~\emph{Organic Architecture} that saw elements of the house as human organs that are in coordination with each other.

In line with the transition to industrialized fabrication, Konrad Wachsmann (May 16, 1901 – November 25, 1980) took the machines as expressions for automation with feedback all the way from products to materials and proposed a universal building method that integrates the designer, the architect, and the construction engineers~\citep{wachsmann1961turning}.

\begin{figure}
  \centering
  \begin{minipage}{.5\textwidth}
    \centering
    \includegraphics[height=4cm]{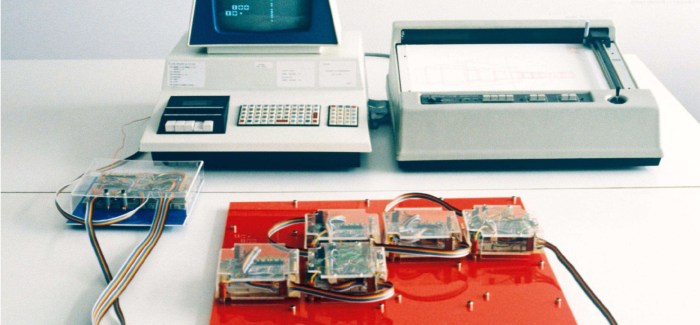}
    \caption{John Frazer's Generator Project\protect\footnotemark}
    \label{fig:generator_project}
  \end{minipage}%
  \begin{minipage}{.5\textwidth}
    \centering
    \includegraphics[height=4cm]{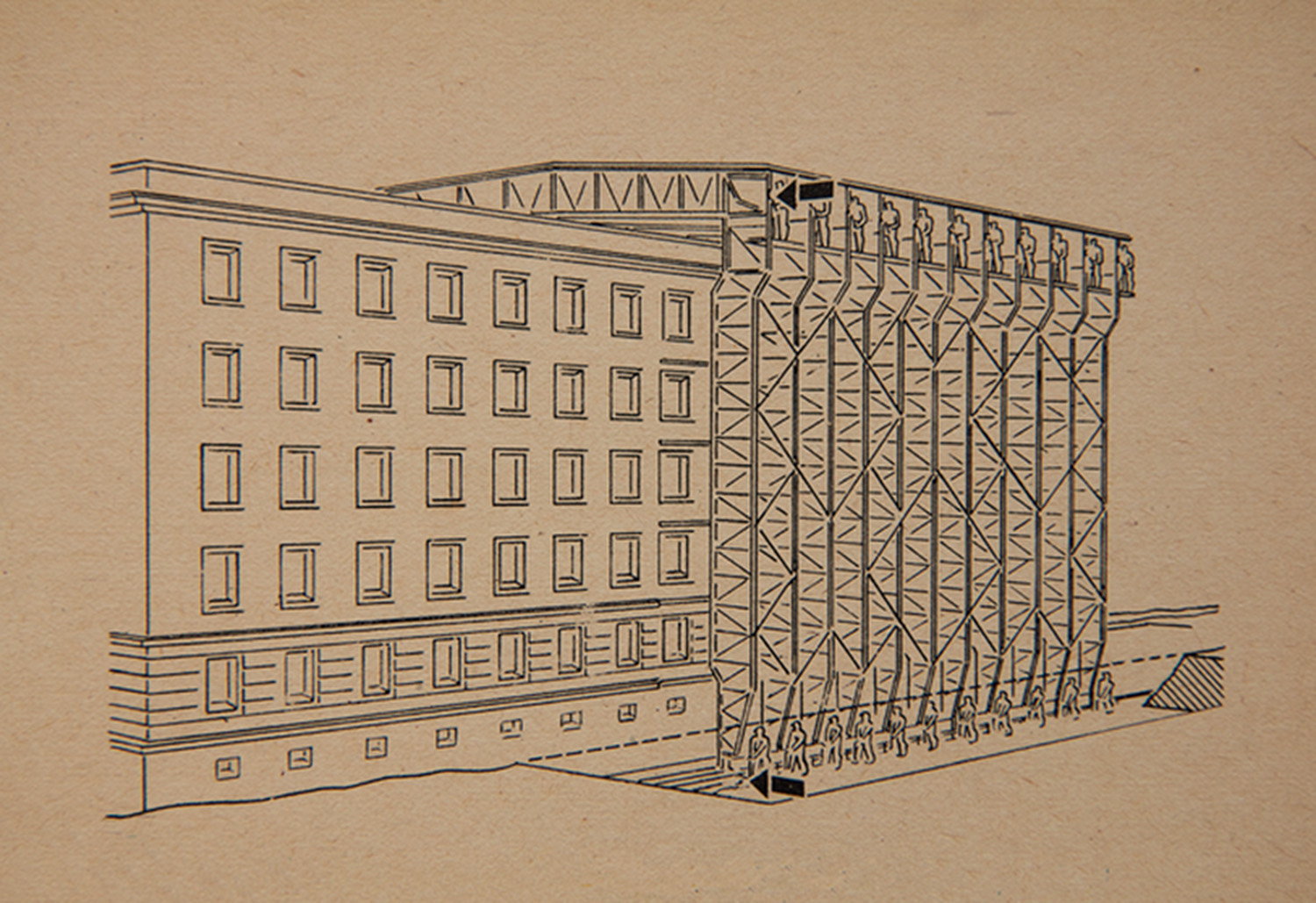}
    \caption{Ernst Neufert's Husbaumaschine Project\protect\footnotemark}
    \label{fig:hausbaumaschine}
  \end{minipage}
\end{figure}

Considering the important role of humans as the residents of houses in the design of buildings, Cedric Price (11 September 1934 – 10 August 2003) proposed interactive buildings that reconfigure in response to the behavior of their users. One futuristic idea was implemented by computer pioneer John Frazer in~\emph{The Generator Project} where the building was assumed as an electrical circuit whose elements relocate according to a written code and within a feedback loop that integrates the behavior of the user and the environment -- see~\Cref{fig:generator_project}. In summary, John Frazer considered architecture as a form of artificial life that evolved in coordination with its surrounding environment\citep{frazer1995evolutionary}.

The development of 3D animation software in the 90s and early 2000s decade allowed architects to visualize their visionary ideas. I can exemplify ``Embryological House'' by Greg Lynn (1964,) and complex design forms by Michael Hansmeyer (1973,) which were later 3D printed to become real objects. The fast progress in the development of 3D design softwares gave architects the unique opportunity to come up with extremely complex designs. However, the implementation of the designs as real-world objects was not straightforward.

\footnotetext{Photos from~\url{https://www.mak.at/}}

The serialized automation line started in 1913 by the Ford assembly line. It took almost thirty years that such serialized automation paved its way to architecture in a project called ``Hausbaumaschine'' by a German architect Ernst Neufert (March 1900 – February 1986) who proposed the standardization of the construction process of buildings -- see~\Cref{fig:hausbaumaschine}. Such a process was highly needed to contribute to the speed and efficiency of rebuilding the German cities during and after World War II.

The robots were introduced to the serial industrial production in the 60s by Ford and General Motors. It took twenty more years when the first attempts were made to use robots for building houses in Japanese company Goyo and Penta-Ocean Construction in 1998 in Tokyo.

Nowadays, robots are used in many stages of the construction process. In recent advances, robotic technology is used to bridge the gap between the abstract 3D computer designs and real-world buildings. In the next section, I'll focus on a couple of technologies developed in one of the leading research groups in Europe that introduce state-of-the-art robotic technology to different stages of the architectural design and construction. Most of the described projects in the next section are chosen from a talk given by Dr. Hannes Mayer in the seminar ``Images of an Artificial'' in Fall 2019 at ETH Z\"urich.

\section{Intelligent Robotics Meet Architecture Research}
\label{sec:current_research}

To exemplify uses of robotics and artificial intelligence in architecture, I discuss some of the technologies developed in Gramazio Kohler Research group from ETH Zürich as one of the world's leading groups in using robotic and computational methods in architecture.

Gramazio Kohler Research group started a branch in digital fabrication in 2005 and chose industrial robots as a universal tool to~\emph{externalize} digital design into the real world. The initial projects done by this group, even though very simple, were good illustrations of the philosophy they follow. For example, ``The Programmed Wall'' is a robot-constructed wall by stacking the bricks with a programmed code that controls the location and angle of the bricks\footnote{\url{https://gramaziokohler.arch.ethz.ch/web/e/lehre/81.html}} . Such a simple program, when implemented by the robot on real material, gives rise to complicated-looking walls whose designs were informed by the computer programs provided by the designer.

Another\addtocounter{footnote}{-2} 
\stepcounter{footnote}\footnotetext{Photo from~\url{http://www.interactivearchitecture.org/the-generator-project.html}}\stepcounter{footnote}\footnotetext{Photo from~\url{https://thefunambulist.net/architectural-projects}}
example of their illustrative work is a robot that pours sand on specified locations for a given period of time\footnote{\url{https://gramaziokohler.arch.ethz.ch/web/e/projekte/321.html}}. Depending on a simple code that only controls how long the sand valve remains open at each location, various-shaped structures are made. The important thing to notice is that the result is not merely caused by the robot or by the material. The combination of the robot's programmed movement and the natural behavior of the material, such as granularity and solidness causes the final outcome of the digital process.

The next project from the same group is a digitally designed wooden pavilion in Rome where a natural hygroscopic property of wood, i.e., ``the wood gets shrunk when dried and gets expanded when it is moistened'', was used to connect the timbers all over the structure without any fastener. A similar idea called Holzdübel-Verbindungen has been practiced in German houses for joints in timber pillars since the $15^\mathrm{th}$ century. The same idea combined with computational design methods allows scaling up the design from simple joints to the entire structure\footnote{\url{https://ethz.ch/en/news-and-events/eth-news/news/2018/08/programming-for-perfect-shade.html}}.

Another example in the same line of projects is based on the behavior of stones as another natural material. The other material used in this project is a sort of string as a man-made material. The string is used to jam stones in order to form stable structures. The idea is to use computational methods that are informed by the behavior of the involved materials to cleverly combine them such that the final outcome takes the desired shape and remains stable under pressure. A robotic arm is used to place the string in a circular shape layer by layer that acts as a belt holding stones together. The robustness of the structure is tested by pressuring it from different angles. Surprisingly, such simple materials gave rise to a stable structure with reasonable resilience analogous to conventional walls where stones are glued together by cement or concrete. The other interesting feature of this structure is the convenience of destruction. Collecting the string from the top releases stones gradually and the entire structure becomes a pile of stones when the string is fully collected. Vertical and horizontal working robots allow complex shapes of structures formed by the same simple law. Even though the current instances of this design is mainly used in art galleries, it can be used in practice as pillars for actual buildings. To showcase its practical uses to the public, the group built several pillars using this method and put a roof on top of them in a public park in Z\"urich and showed that the pillars successfully held the burden for one month as good as steel pillars. One stark difference with traditional architecture is that the outcome of this design is not predicted in advance despite the conventional design methods when the strength of materials in different configurations are all calculated. Here, the synergic effect of all small stones gives rise to the stable pillar whose exact simulation is challenging even using current powerful computers\footnote{\url{https://gramaziokohler.arch.ethz.ch/web/d/projekte/297.html}}.

The important takeaway message from these projects is that by the use of digital computational methods, it is not always necessary to work with fully artificial materials such as carbon fiber. It is possible to use a mixture of fabricated and natural material when the behavior of the natural material is a key factor in the desired outcome.

The combination of natural sciences, digital calculations, and robotics are shown in another project by the same group where the purpose is to stack large stones to form a wall without using any glue or string~\citep{furrer2017autonomous}. Each stone is scanned and its center of gravity is estimated by the computer. Then, its target position is calculated in relation to other stones and an autonomous excavator places the stone accurately in its calculated target location\footnote{\url{https://www.youtube.com/watch?v=bXz52KMGUng}}.

In another project, using Augmented Reality (AR), a recently popularized technology, the group showed the application of AR in the construction process. The design is first calculated by the computer. The bricks are scanned by a camera and their desired locations are calculated according to the target design. A frame of the brick in its target location will appear on the real scene via augmented reality eyeglasses. This allows the worker to know exactly where the brick must be placed\footnote{\url{https://gramaziokohler.arch.ethz.ch/web/e/projekte/371.html}}.

The question of whether the entire process of stacking bricks can become autonomous was one of the initial concerns of the pioneers of artificial intelligence such as Marvin Minsky (August 9, 1927 – January 24, 2016) as he mentioned in his book ``The Society of Mind'', where he emphasized that the construction process requires much more than simple calculus. Due to the multitude of skills necessary to stack bricks to form a proper wall, he connected such capability to general intelligence.

Minsky's argument adds another dimension to the concept of~\emph{artificial} which is~\emph{mind}. Theoretical Philosopher Norman Sieroka (1974,) listed three prominent opposites to~\emph{nature} as Mind (mind vs matter), Technology (man vs robot), and Culture (nature vs nurture).

The last comment I would like to make on this line of projects is the role of~\emph{learning} in architectural processes. In the above-mentioned works, the employed robots are fully programmed and are considered blind in this sense, i.e., the robots do not learn from their experiences. They merely do a simulation followed by execution.

In another project, deep reinforcement learning is used to give learning capability to the robot to fit two elements in each other. A reward function is designed that punishes the robot whenever it fails to complete the task. The algorithm is trained first in simulation and then transferred to reality when the robot is only equipped with a force/torque sensor. The employed algorithm is Deep Deterministic Policy Gradient developed in DeepMind~\citep{lillicrap2015continuous}. It has a generative algorithm that runs the robot and a critique that measures how good the robot is at accomplishing the task. Reinforcement Learning algorithms are easy to think of in architecture since the goal is clear. However, they are hard to implement due to the enormous number of trials they may need to arrive at a good policy to do the task\footnote{\url{https://gramaziokohler.arch.ethz.ch/web/e/forschung/369.html}}.

The combination of all the above-mentioned projects can be distilled into a term ``Robot Builds a House''. This term, hypothesized a robot that takes a computer program as input and manipulate the provided materials to build a house as the output product. This definition abstracts away the construction details and focuses on the role of the robot as a function that maps a computer program into a physical object -- a house in this case.

\section{Reflections}
In this section, I will extend the ideas mentioned in previous sections to areas at the intersection of architecture, intelligent robotics, and algorithmic information theory. First I will take a closer look at the concept of ``Naturalness'' that was introduced in Section~\ref{sec:architecture_history}. Then, using the concept of ``Robot Builds a House'' which implicitly exists in the projects mentioned in Section~\ref{sec:current_research}, I will discuss the potential safety concerns that are probable if the technology is widely deployed. Finally, in a more abstract discussion, I will combine the idea of ``Robot Builds a House'' with Universal Turing Machine to provide a measure of complexity for physical objects such as buildings. As an application, this measure of complexity is used to define a quantifiable measure of beauty.

\subsection{Measure of Naturality}
\label{sec:measure_of_naturality}
Regarding Section~\ref{sec:architecture_history} that was about the evolution of the concept of artificial buildings in history, I can summarize the changes as the evolution of the concept of~\emph{natural}. The more advanced our technology becomes, the more distant the concept of~\emph{natural} will be from the true nature. For example, although caves were considered as natural residential places in the early era, the buildings made of tree branches or even cages are considered as natural buildings in our today's concept of~\emph{natural}.
I would like to provide a different view to the comparison of~\emph{artificial} vs~\emph{natural} things including buildings. I put more emphasis on the appearance of the object than on how it is constructed. To simplify the problem, we can focus only on the photos taken from objects and identify the features of the image that categorize it as natural or artificial. When we look at a scene, there are features that we can objectively see as artificial~\citep{ruderman1994statistics}. I summarize all these features as~\emph{regularity}. Notice that~\emph{regularity} in artificial scenes does not imply that natural scenes are random. Nature imposes its own regularity even though it is totally different from artificial scenes. For example, an extremely straight line in a scene that belongs to a steel product or a carefully smoothed wall made of bricks is a unique character of artificial buildings which are in contrast with those made by nature that often comes with rough and noisy surfaces. Similarly, highly symmetric structures such as circles or curves are other examples of features that discriminate an artificial product from a natural one. Hence, in this definition, a building can be considered natural as long as the unique visual features of artificial buildings are absent.

\subsection{Artificial Nature}
In this section, I consider the question of whether it is possible to construct a building that satisfies the characteristic features of natural buildings as measured by the features proposed in Section~\ref{sec:measure_of_naturality}. To this end, we must first recognize the unique features of a natural scene that rarely occurs in scenes consisting of artificial products. Fractals are one of the stereotypical features of nature that are seen very often in photos taken from natural products~\citep{voss1988fractals}. In abstract terms, a fractal structure refers to a pattern that consists of copies of the same pattern with different scales. Such a pattern is seen in many natural objects such as leaves, roots, branches, snowflakes, broccoli or aerial photos of coastlines, mountains, and glaciers. Such self-similarity seems to be a characteristic feature of nature that is absent in goal-directed industrial products. For example, if the goal of an industrial process is to build a wall with minimum energy and material, embedding a form of such self-similarity will not be advised by engineers. Hence, if one wants to construct a building that looks natural, self-similarity is one of the key features that must be included in every scale. However, current industrial mechanisms are not designed to implement nature-inspired structures. One way to implement self-similarity is to copy the same pattern over and over in the construction process. However, I believe the correct way to build an artificially natural building is to take one step back from the final product and imitate the construction process rather than only the final product. Self-similarity is the outcome of a hierarchical construction process consisting of employing the same process on multiple levels. Hence, the efficient implementation of a nature-inspired building requires implementing the nature-construction process in the first place.

\subsection{Saftey Concerns}
\label{sec:safety}
One of the projects mentioned in~\cref{sec:current_research} was about combining the pre-programmed robots and simple materials such as stones and strings to construct structures that can be used as elements of more complex structures such as buildings. The researchers had tested the robustness of such structures against external disturbances by loading the structures with different loads. Even though the structures built this way, for example, pillars, showed a fair amount of robustness, the type of loads and disturbances were not rich enough to validate the stability of the structure in every condition. Here, I will introduce the concept of~\emph{adversarial robustness} and argue that such a notion of stability can be hazardous in robot-built structures even more than human-built ones.
When a team of humans builds a house, the algorithm they are following is only approximately determined by the architect via a plan. However, construction workers can take minor initiatives at every stage of the construction process which is seen as a source of randomness that makes every building unique in terms of details. This might not be important in a small apartment, but when it comes to large architectural products such as a mall or an opera house, these small details accumulate and give rise to a unique character. However, when a pre-preprogrammed robot is employed to build a building, the used program contains every single movement of the robot. Meaning that, if an adversary has access to this program, it can simulate the final product in a test condition. This transparency gives the adversary all the tools to detect the weaknesses of the building and take advantage of them. For example, if there is a weakness in the construction process that is caused by a programming fault, the weakness is replicated in every building constructed by the same program and the adversary can attack all buildings with the same adversarial plan. However, in a human construction process, the randomness caused by human workers makes it difficult for the adversary to come up with a single attack that applies to all buildings. Notice that this inherent risk is caused by the fact that all the construction process comes from a few lines of deterministic codes that are executed by the robot rather than the complex thinking and decision-making process of human workers.

\subsection{Quantifiable Complexity}
\label{sec:quantifiable_compplexity}
In this section, I define the concept of~\emph{complexity} for buildings as physical objects and show how the conceptual technology ``Robot Builds a House'' can be used to measure this seemingly intractable quantity in practice. Inspired by the algorithmic information theory, the notion of~\emph{Kolmogorov complexity}~\citep{kolmogorov1963tables} is proposed as a measure of complexity for objects. In simple terms, the Kolmogorov complexity of object $x$ is the minimal description of that object. In algorithmic information theory, the description is defined as the computer program $P(x)$ that instructs a general computing machine on how to produce $x$. The general computing machine is known as the Turing Machine~\citep{turing1948intelligent}. A Turing Machine is capable to perform any computation in finite time. In our case, the Turing Machine $T$ takes the description of the object $P(x)$ as input and produces $x$ in a finite time. There may be many programs that all produce $x$ when they are fed to the Turing Machine. Among all programs that produce $x$, the complexity of $x$ is defined as the length of the shortest program. More specifically, the complexity of the object $x$ is defined as $c(x) = \min \{l(x): l(x) \text{ is the length of the computer program } P(x)\text{ that produces the object } x\}$.
An interesting property of Kolmogorov complexity is its independence of the Turing Machine that is used to interpret $P(x)$. To give an intuition on why this is the case, we can restrict ourselves to a class of computing devices called Universal Turing Machines (UTM). UTMs are Turing Machines with the added property that they can reproduce any other Turing Machine. Therefore, a special choice of a Turing Machine appears as an overhead program that remains the same for every input program and makes no difference in the relative complexity of objects since the overhead program is canceled out.

In the following, I establish the connection between the Universal Turing Machine and the ``Robot Builds a House'' technology. The employed robot is considered as a universal machine that is programmed to build something. In this definition, we can see the program defined in the syntax of the programming language of the robot as input $P(x)$. The robot takes the program and produces the house $x$ as its output. Hence, among all programs $\{P_i(x):i=1,2,\ldots,M\}$ that build an identical house, the length of the shortest program is an approximation to the complexity of the house. As mentioned above, this measure of complexity is independent of the robot being used. For convenience, we restrict ourselves to the specific robot which is used to build the houses. Conditioned on a particular robot, we can compare constructed houses in terms of their relative complexity since the complexity of the robot with respect to the universal robot is canceled out in relative comparisons according to the argument I made for Universal Turing Machines. Having a measure of complexity, we can think of several applications. In the following two applications are discussed.

\subsubsection{Quantifiable Measure of Beauty}
\label{sec:measure_of_beauty}
There have been several efforts throughout the history of art towards coming up with a precise definition of beauty. A fundamental question is whether there exists a quantifiable measure of beauty? Inspired by~\citep{schmidhuber1997low}, I propose a quantifiable measure of beauty for the buildings as objects produced by computer programs using the technology ``Robot Builds a House''. Our everyday experience suggests that beauty is a relative term and its perception differs across individuals. This means that the beauty of an object $x$ must be defined as a function of how it is perceived by a particular preceptor. Humans perceive external stimuli via their brains. Let $B(x)$ be the function of the brain that takes the object $x$ and generates the feeling of the beauty of $x$. It is normally assumed that evolution favors energy-efficient systems. Hence, I reasonably assume that $B(x)$ is inversely proportional to the energy consumed by the brain to perceive $x$. Due to the personalized attitude to beauty, I further assume that it is more convenient for the brain $B$ to perceive a set of $N$ patterns $\Theta=\{\theta_1, \theta_2,\ldots, \theta_N\}$. These patterns can be seen as the characteristic parameters of the brain that is now denoted by $B(\cdot;\Theta)$. Hence, if the pattern $x$ is close to one of $\theta_i\in\Theta$, the brain can perceive it with less energy consumption that corresponds with a higher sense of beauty. Notice that $\Theta$ is individual-dependent, i.e., every brain has its own set of easily-perceived patterns. In other words, every brain perceives a particular set of patterns as beautiful. Furthermore, the patterns in $\Theta$ are primitives. I assume the combination of these primitive patterns gives rise to a space spanned by $\Theta$ every member of which is sensed as beautiful by the brain $B(\cdot;\Theta)$.

I argue that a universal measure of beauty can be defined as a sum of two factors: The complexity of the preceptor + the residual complexity of the object that remains out of the space characterized by $\Theta$. To give an intuitive sense of this definition, assume an imaginary brain that perceives every possible pattern. This brain cannot discriminate between beautiful and non-beautiful patterns. On the other end of the spectrum, imagine a brain that is fully blind to the stimuli. This brain can also not discriminate patterns in the input. Hence, we need a fairly simple brain that equally perceives~\emph{a few} patterns~\emph{not too many} patterns. This suggests that the complexity of the brain itself must be taken into account when we talk about the beauty of an object such as a building. Hence, a simple definition of the beauty of $x$ can be $\mathcal{D}(x; \Theta)|\Theta| + r(x|\Theta)$, where $r$ shows how much detail of $x$ is left unexplained by the patters $\Theta$ embedded in the brain. $\mathcal{D}$ is the description length of $x$ given the brain with patterns $\Theta$ and $|\Theta|$ is simply the number of patterns perceived by the brain which is an indication of the complexity of the brain itself. In the next section, I discuss how this definition and assumptions, can be used to construct personalized buildings.

\subsubsection{Optimizing for the Minimal Design}
\label{sec:optimize_for_beauty}
In this section, I discuss the potential idea of designing a house customized to the personalized taste of beauty. I assumed in Section~\ref{sec:measure_of_beauty} that every brain has its own prior patterns based on which it gives beauty score to external patterns. Based on the hierarchical nature of the processing in the brain~\citep{meunier2009hierarchical}, rather than considering the patterns as complete buildings, it is likely that a brain $B$ has a set of primitives denoted by $\Theta_B$ which are combined together to make a complete building. In this sense, infinitely many buildings can be made by a continuous combination of finitely many primitives. Assume the set of buildings made of these primitives is denoted by $\mathcal{S}(\Theta_B)$. The set $\mathcal{S}(\Theta_B)$ consists of all buildings known as beautiful by the brain $B$. However, these primitives concern only the aesthetic aspects of the buildings. Nevertheless, a building must be functional in the first place. Assume a set of constraints $\mathcal{C}(\Theta_B)=\{c_1(\Theta_B),c_2(\Theta_B),\ldots,c_M(\Theta_B)\}$ that must be satisfied by the building primitives to make a functional building. Therefore, the building $x$ which is made by the robot $T$ must be as close as possible to the set of beautifully perceived buildings for the brain $B$ and at the same time satisfy the engineering constraints $\mathcal{C}(\Theta_B)$. This gives us the following optimization criterion 
\begin{equation}
 \label{eq:beauty_criterion}
 P(x) = \mathrm{argmin}_P d(T(P(x), \mathrm{span}(\Theta_B)) + \sum_{i=1}^{i=M}\ell_i(c_i(\Theta_B))
\end{equation}
where $d(\cdot, \cdot)$ is the distance function that determines how far the resultant building $T(P(x))$ is from the space of buildings perceived as beautiful by the brain $B$. The function $\ell_i$ is the cost incurred by violating the $i^\mathrm{th}$ constraint. Solving~\ref{eq:beauty_criterion} gives the program $P$ that when realized by the robot $T$ gives the house $T(P(x))$ that is close to the set of beautiful houses for brain $B$ and also satisfies the engineering constraints $\mathcal{C}$. The distance $d$ between the constructed house and the set of houses constructed using primitives is measured in the sense of Kolmogorov Complexity. Roughly speaking, the distance is the minimal length of the program that constructs those parts of the building which cannot be constructed using the primitives $\Theta_B$. Intuitively, the first term makes sure that the brain can perceive the appearance of the building with minimum energy consumption and hence saw it as a beautiful object. The second term makes sure that the building is designed such that the final product is still functional.

\section{Conclusion}
\label{sec:conclusion}
I used the notion of universality in the concept of ``Robot Builds a House'', that is a universal robot is seen as a generic building machine, to make a connection with Universal Turing Machines as general computing devices. Borrowing concepts from algorithmic information theory, I proposed a method to quantify the complexity of the buildings. Motivated by the hypothesis that human brains have evolved to favor the stimuli which are processed with less energy, I proposed a quantifiable measure of beauty that can be possibly used to customize buildings according to individuals' taste of beauty.

\bibliography{main}

\begin{thebibliography}{11}
\providecommand{\natexlab}[1]{#1}
\providecommand{\url}[1]{\texttt{#1}}
\expandafter\ifx\csname urlstyle\endcsname\relax
  \providecommand{\doi}[1]{doi: #1}\else
  \providecommand{\doi}{doi: \begingroup \urlstyle{rm}\Url}\fi

\bibitem[Emerson and Cameron(1940)]{emerson1940nature}
R.~W. Emerson and K.~W. Cameron.
\newblock \emph{Nature (1836)}.
\newblock Scholars' facsimiles \& reprints, 1940.

\bibitem[Frazer(1995)]{frazer1995evolutionary}
J.~Frazer.
\newblock An evolutionary architecture.
\newblock 1995.

\bibitem[Furrer et~al.(2017)Furrer, Wermelinger, Yoshida, Gramazio, Kohler,
  Siegwart, and Hutter]{furrer2017autonomous}
F.~Furrer, M.~Wermelinger, H.~Yoshida, F.~Gramazio, M.~Kohler, R.~Siegwart, and
  M.~Hutter.
\newblock Autonomous robotic stone stacking with online next best object target
  pose planning.
\newblock In \emph{2017 IEEE International Conference on Robotics and
  Automation (ICRA)}, pages 2350--2356. IEEE, 2017.

\bibitem[Kolmogorov(1963)]{kolmogorov1963tables}
A.~N. Kolmogorov.
\newblock On tables of random numbers.
\newblock \emph{Sankhy{\=a}: The Indian Journal of Statistics, Series A}, pages
  369--376, 1963.

\bibitem[Lillicrap et~al.(2015)Lillicrap, Hunt, Pritzel, Heess, Erez, Tassa,
  Silver, and Wierstra]{lillicrap2015continuous}
T.~P. Lillicrap, J.~J. Hunt, A.~Pritzel, N.~Heess, T.~Erez, Y.~Tassa,
  D.~Silver, and D.~Wierstra.
\newblock Continuous control with deep reinforcement learning.
\newblock \emph{arXiv preprint arXiv:1509.02971}, 2015.

\bibitem[Meunier et~al.(2009)Meunier, Lambiotte, Fornito, Ersche, and
  Bullmore]{meunier2009hierarchical}
D.~Meunier, R.~Lambiotte, A.~Fornito, K.~Ersche, and E.~T. Bullmore.
\newblock Hierarchical modularity in human brain functional networks.
\newblock \emph{Frontiers in neuroinformatics}, 3:\penalty0 37, 2009.

\bibitem[Ruderman and Bialek(1994)]{ruderman1994statistics}
D.~L. Ruderman and W.~Bialek.
\newblock Statistics of natural images: Scaling in the woods.
\newblock In \emph{Advances in neural information processing systems}, pages
  551--558, 1994.

\bibitem[Schmidhuber(1997)]{schmidhuber1997low}
J.~Schmidhuber.
\newblock Low-complexity art.
\newblock \emph{Leonardo}, 30\penalty0 (2):\penalty0 97--103, 1997.

\bibitem[Turing(1948)]{turing1948intelligent}
A.~M. Turing.
\newblock Intelligent machinery, 1948.

\bibitem[Voss(1988)]{voss1988fractals}
R.~F. Voss.
\newblock Fractals in nature: from characterization to simulation.
\newblock In \emph{The science of fractal images}, pages 21--70. Springer,
  1988.

\bibitem[Wachsmann(1961)]{wachsmann1961turning}
K.~Wachsmann.
\newblock \emph{The Turning Point of Building: structure and design}.
\newblock Reinhold Pub. Corp., 1961.

\end{thebibliography}

\end{document}